\documentclass[prd,12pt]{revtex4}
\usepackage{graphicx}
\usepackage{latexsym}
\usepackage{amsmath}
\newcommand{\bq}{\begin{equation}}
\newcommand{\eq}{\end{equation}}
\newcommand{\bqn}{\begin{eqnarray}}
\newcommand{\eqn}{\end{eqnarray}}

\newcommand{\lb}{\label}
\begin{document}
\title{Bounded excursion stable gravastars and black holes}
\author{P. Rocha $^{1}$}
\author{A.Y. Miguelote $^{2}$}
\author{R. Chan $^{2}$}
\author{M.F. da Silva $^{3}$}
\author{N.O. Santos $^{4,5,6}$}
\author{Anzhong Wang $^{3,4,7}$}
\email{anzhong_wang@baylor.edu}
\affiliation{$^{1}$ Instituto de F\'{\i}sica, Universidade Federal
Fluminense, Av. Litor\^anea, s/n, Boa Viagem 24210-340, Niter\'oi, RJ,
 Brazil\\
$^{2}$ Coordena\c{c}\~ao de Astronomia e Astrof\'{\i}sica, Observat\'orio
Nacional, Rua General Jos\'e Cristino, 77, S\~ao Crist\'ov\~ao  20921-400, Rio
de Janeiro, RJ, Brazil\\  
$^{3}$ Departamento de F\'{\i}sica Te\'orica, 
Instituto de F\'{\i}sica, Universidade do Estado do Rio de Janeiro, 
Rua S\~ao Francisco Xavier 524, Maracan\~a
20550-900, Rio de Janeiro - RJ, Brasil\\
$^{4}$ LERMA/CNRS-FRE 2460, Universit\'e Pierre et Marie Curie, ERGA, 
Bo\^{\i}te 142, 4 Place Jussieu, 75005 Paris Cedex 05, France\\
$^{5}$ School of Mathematical Sciences, Queen Mary,
University of London, London E1 4NS, UK\\ 
$^{6}$ Laborat\'orio Nacional de Computa\c{c}\~{a}o Cient\'{\i}fica, 
25651-070 Petr\'opolis RJ, Brazil\\
$^{7}$ GCAP-CASPER, Department of Physics,
Baylor University, Waco, TX 76798, USA}

\begin{abstract}
 
Dynamical models of prototype gravastars were constructed in order to 
study their stability. The models are the  Visser-Wiltshire 
three-layer gravastars, in which  an infinitely thin spherical shell 
of stiff fluid divides the whole spacetime into two regions, where 
the internal region is  de Sitter, and  the external is  Schwarzschild. 
It is found that in some cases the  models represent the ``bounded 
excursion"  stable gravastars, where  the thin shell is oscillating 
between two finite radii, while in other cases they collapse until the 
formation of black holes. In the phase space, the region for the ``bounded 
excursion"  gravastars is very small in comparison to that of black holes, 
but not empty. Therefore, although the existence of gravastars cannot be 
excluded from such dynamical models, our results do indicate that, 
even if gravastars indeed exist, they do not exclude the existence of 
black holes.  

\end{abstract}
\pacs{98.80.-k,04.20.Cv,04.70.Dy}
\preprint{arXiv: xxxxxxxx}
\maketitle

\section{Introduction}

\renewcommand{\theequation}{1.\arabic{equation}}
\setcounter{equation}{0}

Black holes are well known and accepted objects  not only on the scientific community 
but also on the general public. This is, for example, attested to partially by the 
phenomenal success of the well-known film, {\em Black Holes} (New River Media, 
1998), directed by Pappi Corsicato, and by several best selling popular
books, such as Kip S. Thorne, {\em Black Holes and Time Warps:
Einstein's Outrageous Legacy} (W.W. Norton $\&$ Company, 1995),
and Stephen Hawking, {\em A Brief History of Time} (Bantam Books
publisher, 10th edition, 1998). However, the real detection of black holes
highly demands their detailed explorations, including the forms of 
gravitational waves emitted by them. In principle, such a detection can never be 
conclusive and is fundamentally impossible \cite{AKL02}, although there are 
many strong astronomical evidences for their existence. 

The simplest example of black holes is the spherically symmetric vacuum solution of the 
Einstein field equations,
\bq
\lb{1.1}
ds^{2} = -f(r) dt^{2} + \frac{1}{f(r)} dr^{2} 
   + r^{2}\left(d\theta^{2} + \sin^{2}\theta d\varphi^{2}\right),
\eq
where the function $f(r)$ is given by
\bq
\lb{1.2}
f(r) =  1 - \frac{2M}{r},
\eq
in the units where $c = 1 = G$. The metric is singular at both $r = 0$ and $r = r_{g}
\equiv 2M$. However, the nature of these singularities is different. In particular, 
the one at $r = 0$ is generic and the spacetime curvature diverges there, while the one
at $r = r_{g}$ is a coordinate one and  can be made disappear after proper coordinate transformations.
A classical point test particle will freely fall through $r = r_{g}$ 
without experiencing anything special. Such a surface is usually called as an event horizon.
The studies of the properties of such horizons are fundamentally important, and it is found 
that there may exist a deep connection between gravity and thermodynamics 
\cite{bardeen}. The discovery of the quantum Hawking radiation \cite{hawking} and the 
black hole entropy which is proportional to the area of the event horizon of the black hole
\cite{bekenstein} further supports this idea. This interesting relation was first manifested
when Jacobson derived the Einstein field  equations from the first law 
of thermodynamics (FLT) by assuming the proportionality of the entropy and the horizon area
for all local acceleration horizons \cite{ted}.  For  static spherically symmetric and stationary 
axisymmetric space-times, Padmanabhan {\em et al}  showed that Einstein's equations at the horizon 
give rise to the FLT \cite{padmanabhan02,dawood}. Such considerations were further generalized to  
the Lovelock gravity  \cite{padmanabhan06}. In \cite{eling}, on the other hand, the gravitational 
field equations (GFE's) for the nonlinear  $f(R)$  theory were derived from the FLT by adding 
some non-equilibrium corrections.

For the de Sitter space there also exist Hawking temperature and entropy associated with the
cosmological event horizons  and its  thermodynamic laws \cite{gibbons}. In this space, 
the event horizons coincides with the apparent horizon. For more general cosmological 
models,  event horizons may not exist, but apparent horizons always do, so it is 
possible to have Hawking temperature and entropy associated with apparent horizons. 
Along this line, the connection between the FLT of apparent horizons and 
the Friedmann equations in Einstein's theory with/without the Gauss-Bonnet 
term, as well as in the Lovelock theory, were found  \cite{cai05}. 
More recently, it was found that this is also true  for the braneworld 
cosmology \cite{ge}.  In  \cite{GW07}, with the help of a new mass-like function, it was  
shown that the FLT of apparent horizons in equilibrium can be derived from the Friedmann 
equations in various theories of gravity, including the Einstein, scalar-tensor, nonlinear 
$f(R)$, and Lovelock.

\begin{figure}
\centering
\includegraphics[width=12cm]{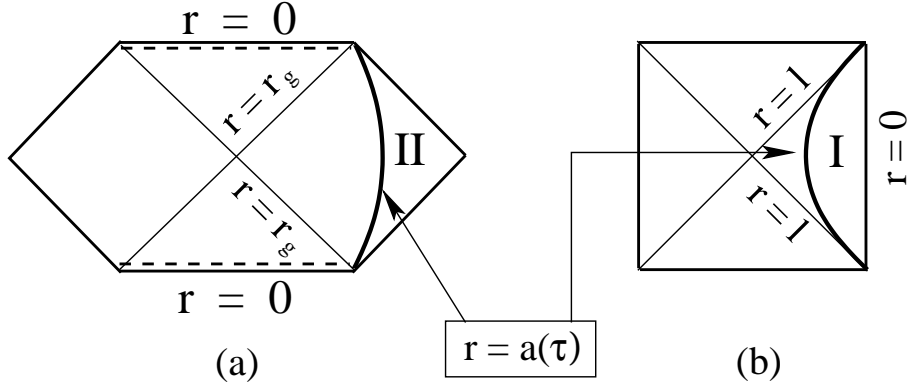}
\caption{(a) The Penrose diagram for the Schwarzschild vacuum solution. Region $II$ is the region
where $r > a(\tau)$.  (b) The Penrose diagram for the de Sitter solution. Region $I$ is the region
where $r < a(\tau)$.  An infinitely thin shell located at   $r = a(\tau)$  connects these two regions
to form a dynamical spacetime of a prototype gravastar, as shown in Fig. \ref{fig2}.} 
\label{fig1}
\end{figure}

\begin{figure}
\centering
\includegraphics[width=12cm]{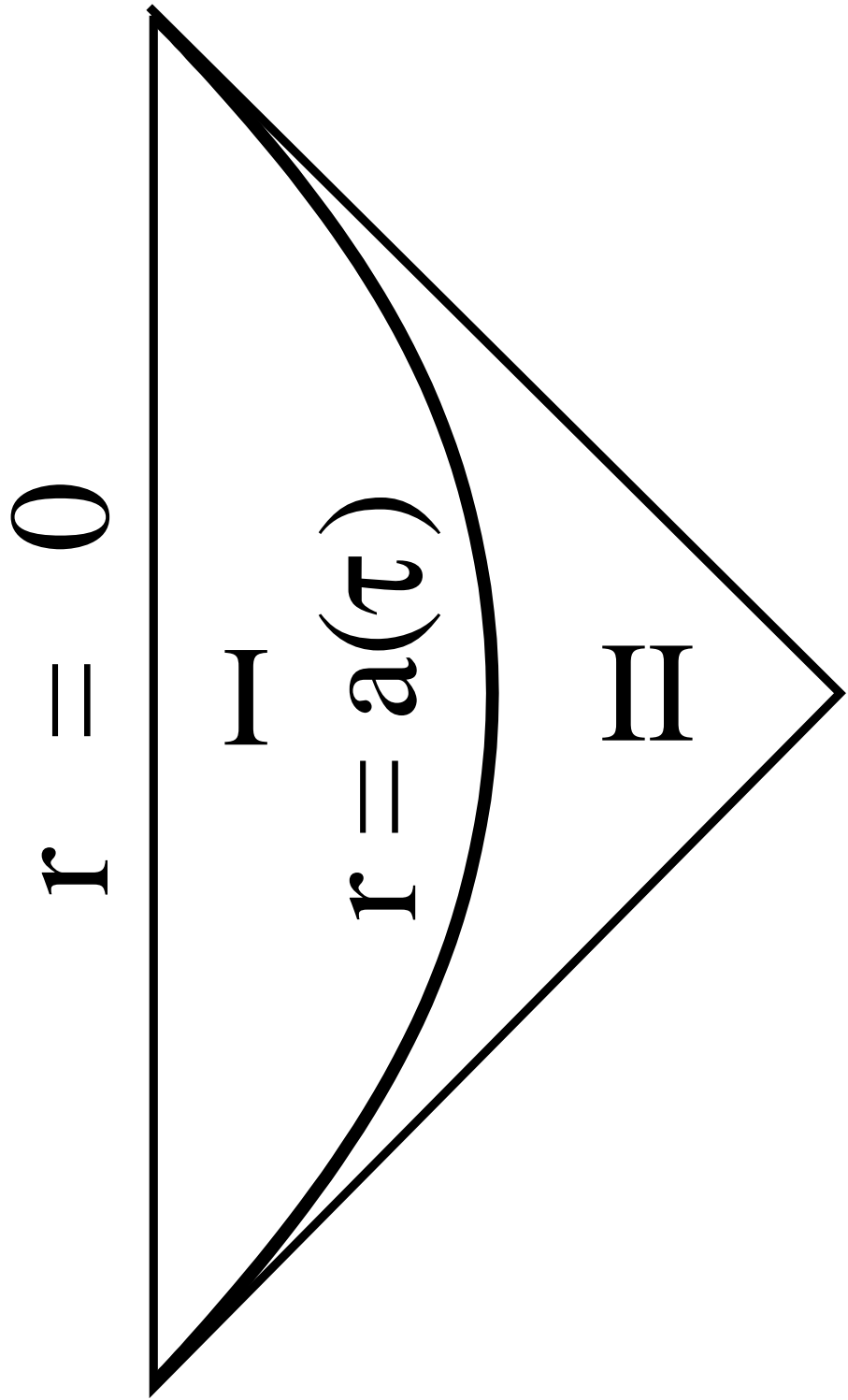}
\caption{The Penrose diagram for the solution given by Eq.(\ref{1.3}). The  infinitely thin shell 
 at   $r = a(\tau)$  connects the two regions $I$ and $II$ shown in Fig. \ref{fig1}
to form a spacetime of a prototype gravastar.} 
\label{fig2}
\end{figure}

Despite of all these theoretical and observational successes, a number of paradoxical 
problems reparging to black holes also exist \cite{Wald01}, which 
frequently motivate authors to look for other alternatives, in which the endpoints of 
gravitational collapse are massive stars without  horizons. Example of such models include
gravastars \cite{MM01}, Bose superfluid \cite{CHLS03}, and black stars \cite{Vach07},
to name only few of them. Among these models,  gravastars have  received particular attention
recently \cite{grava}, partially due to the tight connection between the cosmological constant
and a currently accelerating universe \cite{DEs}, although very strict observational constraints 
on the existence of such stars may exist \cite{BN07}.
In the original model of Mazur and  Mottola (MM) \cite{MM01}, gravastars consist of five layers:
an internal core $0 < r < r_{1}$, described by the de Sitter universe, an intermediate thin layer 
of stiff fluid $r_{1} < r < r_{2}$, an external region $ r > r_{2}$, described by the 
Schwarzschild solution (\ref{1.1}) - (\ref{1.2}), and two infinitely thin shells, appearing,
respectively, on the hypersurfaces $r = r_{1}$ and $r = r_{2}$. By properly choosing
the free parameters involved, one can show that the two shells can have only tensions but with
opposite signs \cite{MM01}. Visser and Wiltshire (VW) argued that such five-layer models can be
simplified to three-layer ones \cite{VW04}, in which the two infinitely thin shells and the 
intermediate region are replaced by one infinitely thin shell, so that the function $f(r)$ 
in the metric (\ref{1.1}) is 
given by
\bq
\lb{1.3}
f(r) = \begin{cases} 1 - \frac{2M}{r}, &  r > a(\tau),\\
1 - \left(\frac{r}{l}\right)^{2}, &  r < a(\tau),
\end{cases}
\eq
where $r = a(\tau)$ is a timelike hypersurface, at which the infinitely thin shell is located,
and $\tau$ denotes the proper time of the thin shell. The constant $l \equiv \sqrt{3/\Lambda}$ 
denotes  the de Sitter radius.
On the hypersurface $r = a(\tau)$ Israel junction conditions yield
\bq
\lb{1.4}
\frac{1}{2}\dot{a}^{2} + V(a) =  0,
\eq
where an overdot denotes the derivative with respect to the proper time $\tau$
of the thin shell. 
Therefore, in the region $ r > a(\tau)$ the spacetime is locally Schwarzschild, while in the
region $ r < a(\tau)$ it is locally de Sitter.  These two different regions are connected through
a dynamical infinitely thin  shell located at $r = a(\tau)$ to form a new spacetime of gravastar
[cf. Figs. \ref{fig1} and \ref{fig2}].

To study the dynamics of Eq.(\ref{1.4}), one can follow two different approaches: one is
to prescribe a potential $V(a)$ and leave the equation of state of the shell as derived, and the
other is to  prescribe an equation of state of the shell and leave the  potential $V(a)$ as derived.
VW followed the first approach, and studied in details the case where
\bq
\lb{1.5}
V\left(a_{0}\right) = 0, \;\;\; V'\left(a_{0}\right) = 0, \;\;\;
V''\left(a_{0}\right) > 0,
\eq
where a prime denotes the ordinary differentiation with respect to the indicated argument.
If and only if there exists such an $a_{0}$ for which the above conditions are satisfied,
the model is said to be stable. 
Among other things, VW found that there are many equations of state for which the gravastar
configurations are stable, while others are not \cite{VW04}. Carter studied  the same
problem and found new equations of state for which the gravastar is stable \cite{Carter05}, 
while De Benedictis {\em et al} \cite{DeB06} and Chirenti and Rezzolla \cite{CR07} 
investigated the stability of the original model
of  Mazur and  Mottola agaist axial-perturbations, and found that gravastars are stable to
these perturbations. Chirenti and Rezzolla also showed that their quasi-normal modes differ from
those of a black hole of the same mass, and thus can be used to discern a gravastar from a
black hole.

As VW noticed, there is a less stringent notion of stability, the so-called ``bounded excursion" 
models, in which there exist two radii $a_{1}$ and $a_{2}$ such that
\bq
\lb{1.6}
V\left(a_{1}\right) = 0, \;\;\; V'\left(a_{1}\right) \le 0, \;\;\;
V\left(a_{2}\right) = 0, \;\;\; V'\left(a_{2}\right) \ge 0,
\eq
with $V(a) < 0$ for $a \in \left(a_{1}, a_{2}\right)$, where $a_{2} > a_{1}$.

In this paper, our purpose is twofold: (a) First, we construct three-layer VW dynamical models,
some of which represent ``bounded excursion" stable gravastars, and some represent the
collapse of a prototype gravastar, where the final fate of the collapse
 is the formation of a black hole.
(2) Second, in the phase space we compare the region of such stable gravastars with
the one of black holes,  and show explicitly that both of them are non-zero, although
the former is much smaller than the latter. The rest of the paper is  organized as follows: 
In Sec. II we shall study various cases, in which
all the possibilities of forming black holes, gravastars, de Sitter, and Minkowski spacetime
exist. In Sec. III we present our main conclusions.

Before turn to the next section, we note some relevant work. In particular, the 
gravitational collapse of dark energy in the background of dark matter was studied in 
\cite{CW06}, and it was found that when only dark energy is present, black holes are  
never formed. When both of them are present, black holes can be formed, 
due to the condensation of the dark matter. Similar results were obtained in \cite{6a,6b,Hara}.
Recently, such studies were further generalized to a homogeneous and isotropic expanding 
Friedmann-Robertson-Walker universe dominated by dark energy \cite{LW07}.

\section{Formation of Gravastars from Gravitational Collapse}

\renewcommand{\theequation}{2.\arabic{equation}}
\setcounter{equation}{0}

To keep the ideas of MM as much as possible, we consider the thin shell as consisting
of a stiff fluid, $\sigma = - \vartheta$, where $\sigma$ and $\vartheta$ denote, respectively,
the surface energy density and tension of the shell. Then, we find that \cite{VW04}
\bq
\lb{2.1a}
\sigma = \sigma_{0}\left(\frac{a_{0}}{a}\right)^{4},
\eq
where $\sigma_{0}$ and $a_{0}$ are integration constants, and have dimensions of
surface energy density and length, respectively. It can be shown that the potential 
 appearing in Eq.(\ref{1.4}) now can be cast in the form,  
\bqn
\lb{2.1}
V(R) = - \frac{1}{2}\left( -1 + \frac{m}{R} + \frac{1}{4R^{6}} + m^{2} R^{4} + \frac{R^{2}}{2L^{2}}
 - \frac{mR^{7}}{L^{2}} + \frac{R^{10}}{4L^{4}}\right),
\eqn
where
\bq
\lb{2.2a}
m \equiv \frac{M}{k^{1/3}},\;\;\;
R \equiv \frac{a}{k^{1/3}}, \;\;\;
L \equiv \frac{l}{k^{1/3}},
\eq
with $k \equiv 4\pi a_{0}^{4}\sigma_{0}$. 
Therefore, for any given constants $m$ and $L$, Eq.(\ref{1.4}) uniquely determines the collapse 
of the prototype  gravastar. Depending on the initial value $R_{0}$,  the collapse can
form either a black hole, or gravastar,  or a Minkowski, or a de Sitter space. 
In the last case, the thin shell
first collapses to a finite non-zero minimal radius and then expends to infinity.  To  guarantee  
that initially the spacetime does not have any kind of horizons,  cosmological or event,
we must restrict $R_{0}$ to the range,
\bq
\lb{2.2b}
2m < R_{0} < L,
\eq
correspondingly $r_{0} \in (r_{g}, l)$. When $m = 0= \Lambda$, the thin shell disappears,
and the whole spacetime is Minkowski. So, in the following we shall not consider this case, and begin
with the one $ m = 0$ and $\Lambda \not= 0$.

\subsection{$m = 0$ and $\Lambda \not= 0$}

\begin{figure}
\centering
\includegraphics[width=12cm]{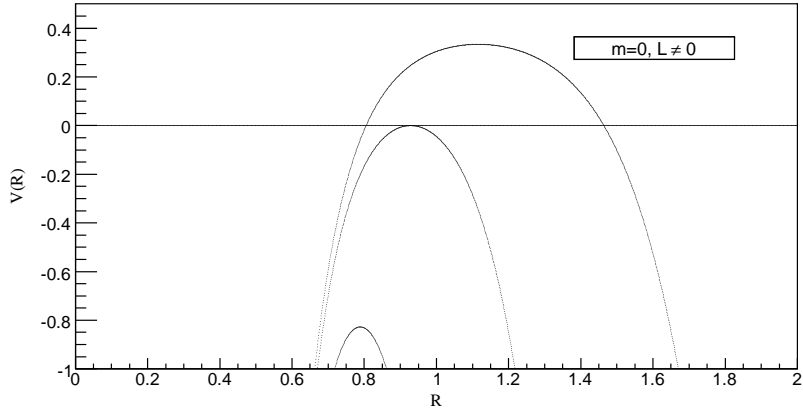}
\caption{The potential $V(R)$ for $m = 0$. The top line is for $L =2.0 > L_{c} \simeq 0.9588$,
the middle line is for $L = L_{c}$, and the bottom line is for $L = 0.5 < L_{c}$. } 
\label{Vm0}
\end{figure}

\begin{figure}
\centering
\includegraphics[width=12cm]{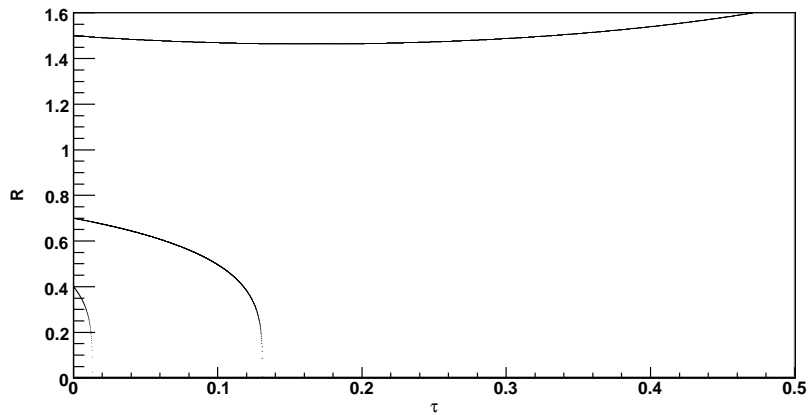}
\caption{The motion of the shell $R(\tau)$ vs the proper time, $\tau$, of the shell for $m = 0$. 
The top line is for $L =2.0 > L_{c} \simeq 0.9588$,
the middle line is for $L = L_{c}$, and the bottom line is for $L = 0.5 < L_{c}$. } 
\label{Rm0}
\end{figure}

In this case, the spacetime outside the thin shell is flat, and the  mass of the shell completely
screens the  mass of the internal de Sitter spacetime. From Eq.(\ref{2.1}) we find that 
\bq
\lb{2.2ab}
V(R) =  \frac{1}{2}\left(1 -   \frac{1}{4R^{6}} - \frac{R^{2}}{2L^{2}}
 -  \frac{R^{10}}{4L^{4}}\right).
\eq
Then, it can be shown that the equations $V(R) = 0$ and $V'(R) = 0$ have
the explicit solution,
\bq
\lb{2.2aa} 
L  = L_{c} \equiv \left[\frac{5}{3}\left(\frac{4}{5}\right)^{8/3}\right]^{1/2} \simeq 0.9588,   \;\;\;
R =  R_{min} \equiv \left(\frac{4}{5}\right)^{1/3}.
\eq
For $L < L_{c}$ the potential $V(R)$  is strictly negative as shown in Fig. \ref{Vm0}. As
a result, if the star starts to collapse at $R = R_{0}$, it will collapse  continuously until $R = 0$, 
whereby a Minkowski spacetime is formed,
as shown  by the bottom line in  Fig. \ref{Rm0}.
When $L = L_{c} \simeq 0.9588$, since $ R_{0} < L_{c}$,  we can see that, similar to the last case,
the star will collapse until the center $R = 0$ and turns the whole spacetime into  Minkowski, 
as shown by the middle line in  Fig. \ref{Rm0}. For $L > L_{c}$, the potential $V(R)$  
is positive between $R_{1}$ and $R_{2}$, where $R_{1,2}$ are the two real roots of the equation 
$V(R, L > L_{c}) = 0$ with $R_{2} > R_{1} > 0$. In this case, if the star starts to collapse with
$R_{0} < R_{1}$, as can be seen from Fig. \ref{Vm0},  it will collapse to $R = 0$,
whereby a Minkowski spacetime is finally formed. If it starts to collapse with
$R_{0} > R_{2}$,  it will first collapse to $R = R_{2}$ and then starts to expand until 
$R = \infty$, and the whole spacetime is finally   de Sitter, as shown by the top 
line in Fig. \ref{Rm0}.

\subsection{$\Lambda = 0 $ and $m \not= 0$}

\begin{figure}
\centering
\includegraphics[width=12cm]{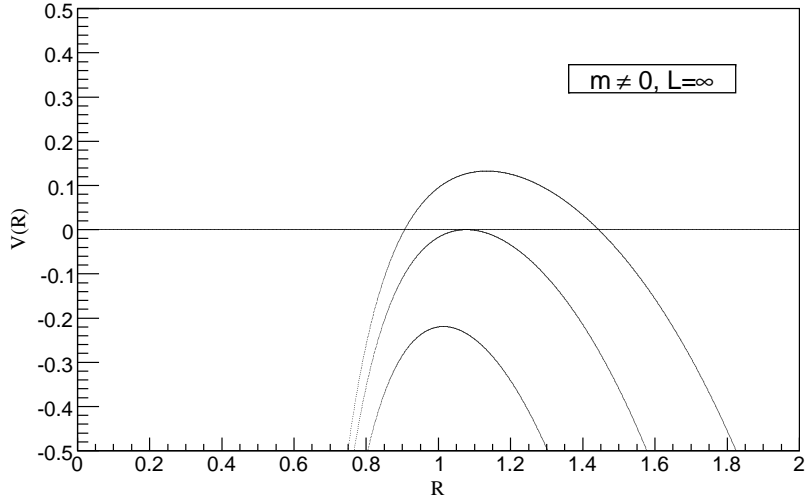}
\caption{The potential $V(R)$ for $\Lambda = 0$ (or $L = \infty$). The top line is for $m < m_{c}
\simeq 10^{-0.286}$, the middle line is for $m = m_{c}$, and the bottom line is for $m > m_{c}$. } 
\label{VLinfty}
\end{figure}

\begin{figure}
\centering
\includegraphics[width=12cm]{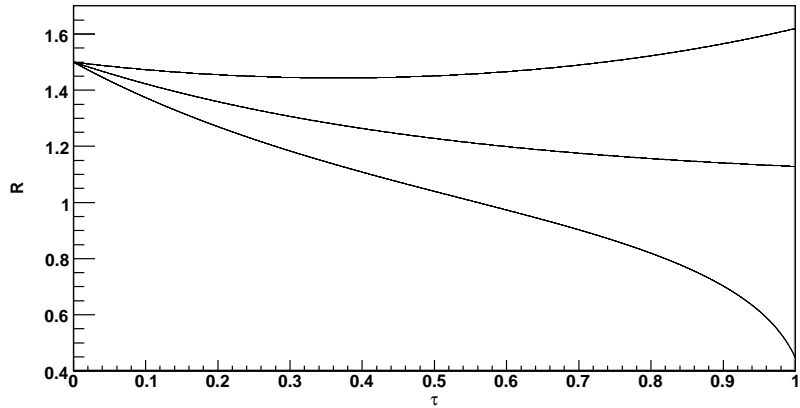}
\caption{The motion of the shell $R(\tau)$ vs its proper time $\tau$ for $\Lambda = 0$. 
The top line is for $m < m_{c} \simeq 10^{-0.286}$,
the middle line is for $m = m_{c}$, and the bottom line is for $m > m_{c}$. } 
\label{RLinfty}
\end{figure}

In this case, Eq.(\ref{2.1}) yields,
\bq
\lb{2.2c}
V(R) =  \frac{1}{2}\left(1 - \frac{m}{R} - \frac{1}{4R^{6}} - m^{2} R^{4}\right),
\eq
from which we find that the equations $V(R) = 0$ and $V'(R) = 0$ have
the explicit solution,
\bq
\lb{2.2} 
m = m_{c} \equiv \frac{3}{4}\left(\frac{4}{5}\right)^{5/3} \simeq 10^{-0.286}, \;\;\;
R = R_{c} \equiv \left(\frac{3}{4m_{c}}\right)^{1/5} = \left(\frac{5}{4}\right)^{1/3}.
\eq
For $m > m_{c}$ the potential $V(R)$  is strictly negative as shown in Fig. \ref{VLinfty}. Then, the 
collapse always forms black holes, as shown clearly by the bottom line in Fig. \ref{RLinfty}.
For $m = m_{c}$, there are two different possibilities, depending on the choice of the
initial radius $R_{0}$. In particular, if the star begins to collapse with $R_{0} > R_{c}$,
the collapse will asymptotically approach the minimal radius $R_{c}$. Once
it collapses to this point, the shell will stop collapsing  and remains there for ever, as can be seen
from  the middle line in Fig. \ref{RLinfty}. However, in this case this point is unstable, 
and any small perturbations will lead the star either to expand for ever and leave behand a flat
spacetime, or to collapse  until $R = 0$, whereby a  Schwarzschild black hole is finally formed. 
On the other hand, if the star begins to collapse with $2m_{c} < R_{0} < R_{c}$, as shown by 
Fig. \ref{VLinfty}, the star will collapse until a black hole is formed. 
For $m < m_{c}$, the potential $V(R)$  
has a positive maximal, and the equation $V(R, m < m_{c}) = 0$  has two positive roots $R_{1,2}$  
with $R_{2} > R_{1} > 0$. As in the last case, now  there are also two possibilities,
depending on the choice of the initial radius $R_{0}$. If $R_{0} > R_{2}$, the star will first
collapse to its minimal radius $R = R_{2}$ and then expand to infinity, whereby a Minkowski
spacetime is finally formed, as shown by the top line in Fig. \ref{RLinfty}. 
If  $2m < R_{0} < R_{1}$,  the star will collapse continuously until $R = 0$, and  a black hole will be
finally formed.

\subsection{$m \not= 0 $ and $\Lambda \not= 0$}
 
In this case, from Eq.(\ref{2.1}) we find that the equations $V(R) = 0$ and $V'(R) = 0$ have
the solution of the form, $m = m_{c}(L)$ for any given $L$. The exact dependence of $m_{c}$ on 
$L$ cannot be given explicitly. Instead, in the following we consider some representative cases.

\subsubsection{$m = 10^{-0.286}, \;\; L = 2.5 \times 10^{5}$}

\begin{figure}
\centering
\includegraphics[width=12cm]{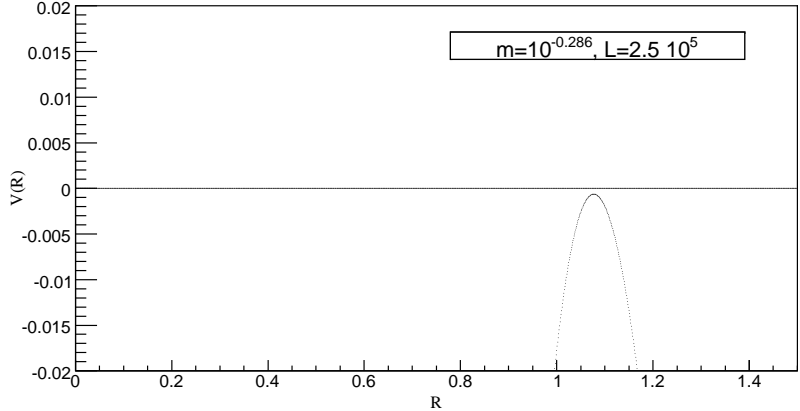}
\caption{The potential $V(R)$ for $ m = 10^{-0.286}$ and $L = 2.5 \times 10^{5}$, 
which is always negative.} 
\label{Vca}
\end{figure}

If we set $m = m_{c} \simeq 10^{-0.286}$ and graduately turn on the cosmological constant,
we find that the potential $V(R)$ becomes completely negative, as shown by Fig. \ref{Vca}.
Then, for any given $R_{0}$ with $2m < R_{0} < L$, the star will always collapse to form
a black hole. Comparing it with the case $m =   m_{c}$ and $\Lambda = 0$, we find that
the presence of the cosmological constant makes the collapse more like to form black holes 
than  gravastars, or any of the others.

\subsubsection{$m = 10^{-4}, \;\; L = 2.5 \times 10^{5}$}

\begin{figure}
\centering
\includegraphics[width=12cm]{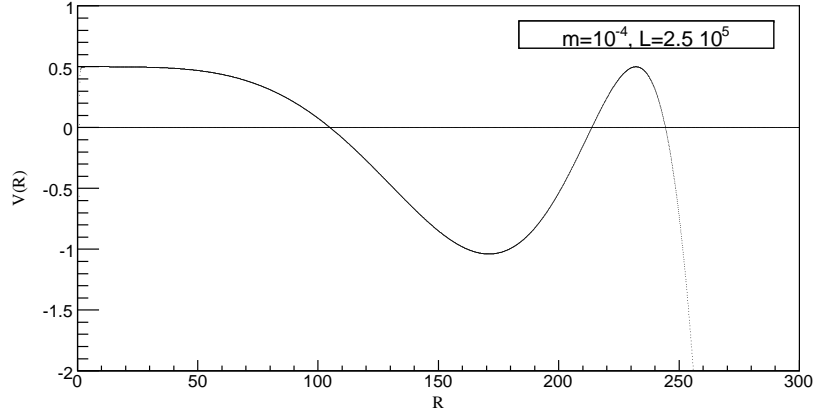}
\caption{The potential $V(R)$ for $ m = 10^{-4}$ and $L = 2.5 \times 10^{5}$. } 
\label{Vcb}
\end{figure}

\begin{figure}
\centering
\includegraphics[width=12cm]{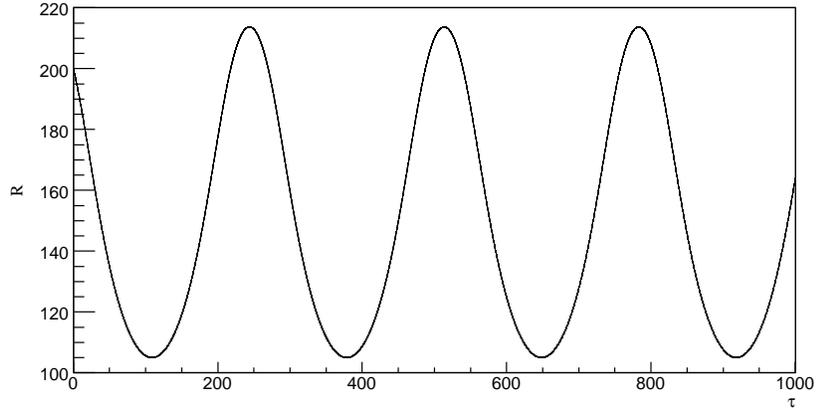}
\caption{The development of $R(\tau)$ vs the proper time $\tau$  for $ m = 10^{-4}$ and 
$L = 2.5 \times 10^{5}$.} 
\label{Rcb}
\end{figure}

\begin{figure}
\centering
\includegraphics[width=12cm]{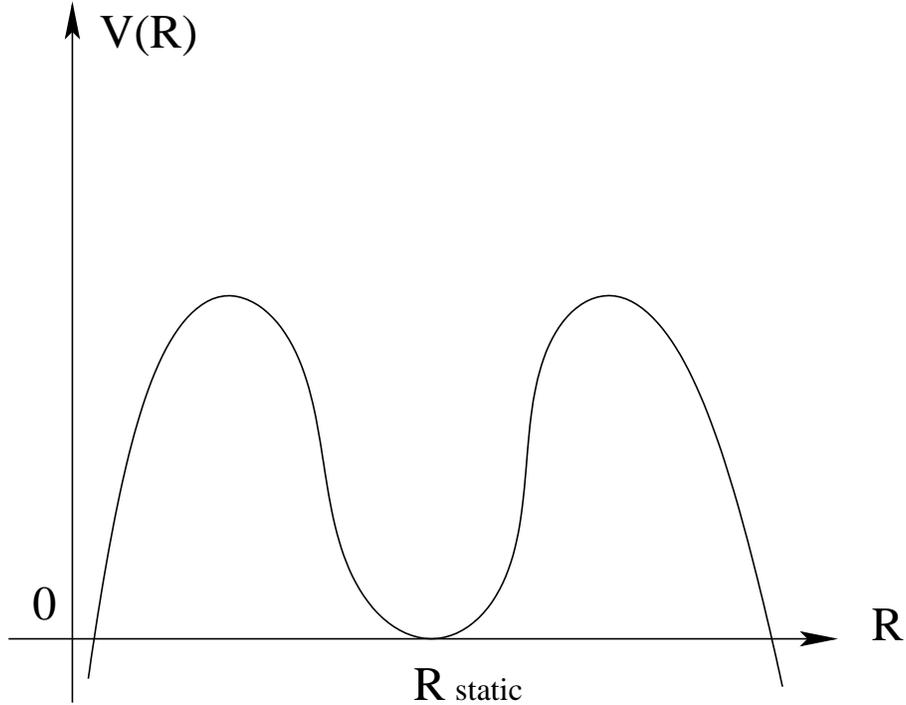}
\caption{The potential $V(R)$ for  the formation of a gravastar in a realistic collapse,
after the star settles down to the minimus point $R = R_{static}$, where 
$V\left(R=R_{static}\right) = 0 = V'\left(R=R_{static}\right)$. } 
\label{fig3}
\end{figure}

If we keep $L$ fixed, i.e., $L = 2.5 \times 10^{5}$, and tun $m$ downward, we find that,
for $m = 10^{-4}$, the potential takes the shape given by Fig. \ref{Vcb}, from which we can see that
$V(R) = 0$ now has four real roots, say, $R_{i}$, where $R_{i+1} > R_{i}$.
If we choose $R_{0} > R_{4}$, then the star will first collapse to
$R = R_{4}$, and then expand to infinity, whereby a de Sitter space is finally formed. However, 
if we choose $R_{2} < R_{0} < R_{3}$, the collapse will bounce back and forth between $R = R_{2}$ 
and $R = R_{3}$. Such a  possibility is shown in Fig. \ref{Rcb}. This is exactly  
the  so-called ``bounded excursion" model mentioned in \cite{VW04}, but was not studied
there or somewhere else.  Of course, in a realistic situation, the star will emit both gravitational waves and 
particles, and the potential shall be self-adjusted to produce a minimum at $R = R_{static}$ where 
$V\left(R=R_{static}\right) = 0 = V'\left(R=R_{static}\right)$
[cf. Fig. \ref{fig3}], whereby a gravastar is finally formed \cite{VW04}.  

\begin{figure}
\centering
\includegraphics[width=12cm]{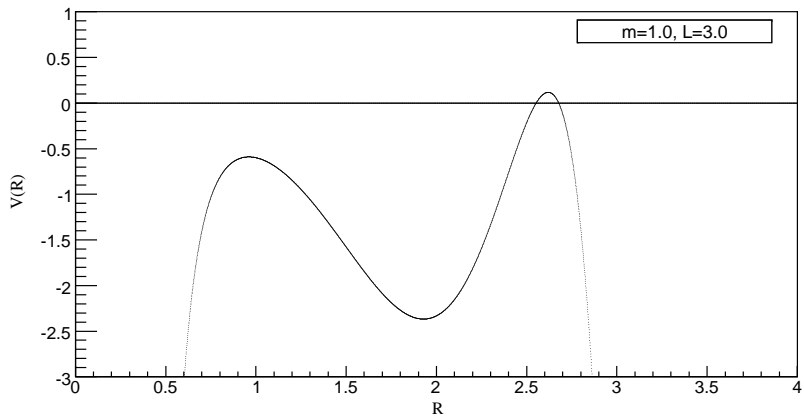}
\caption{The potential $V(R)$ for $ m = 1.0$ and $L = 3.0$. } 
\label{Vcc}
\end{figure}

\begin{figure}
\centering
\includegraphics[width=12cm]{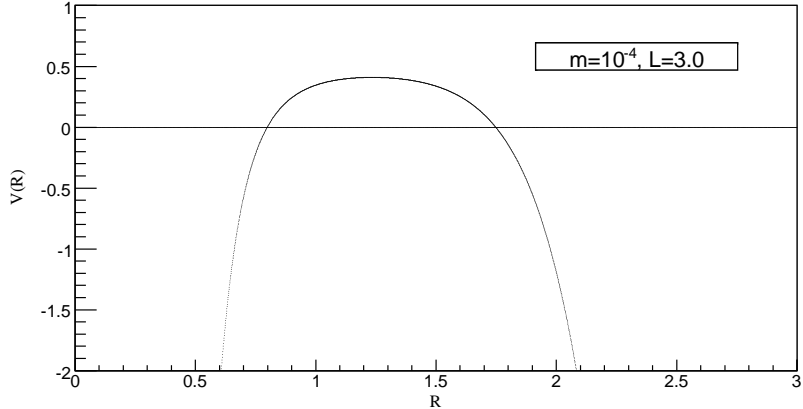}
\caption{The potential $V(R)$ for $ m = 10^{-4}$ and $L = 3.0$. } 
\label{Vcd}
\end{figure}

Although such a shape of potential is very difficult to find in the phase space of $m$ and $L$,
its measurement is not zero. So, one cannot completely exclude the existence
of gravastars. However, we do find that it is
easy to find potentials that lead to the formation of black holes. Figs. \ref{Vcc} and \ref{Vcd}
are for $m = 1.0$ and $L = 3.0$, and $m = 10^{-4}$ and $L = 3.0$, respectively. From these
figures we can see that, by properly choosing the initial radius $R_{0}$, the collapse always 
forms black holes. In  contrary, in these cases gravastars cannot be formed no matter how to tun 
$R_{0}$. On the other hand, we are not able to find values of $m$ and $L$ for which only gravastars 
can be formed. 

\section{Conclusions}

In this paper, we have studied the problem of the stability of gravastars by
constructing dynamical three-layer models  of VW \cite{VW04},   
which consists of an internal de Sitter space, a dynamical infinitely thin  shell of 
stiff fluid, and an external Schwarzschild spacetime. 
We have shown explicitly that the final output can be a black
hole, a   ``bounded excursion" stable gravastar, a Minkowski, or a de Sitter spacetime,
depending on the total mass $m$ of the system,  the cosmological constant $\Lambda$, and  
the initial position $R_{0}$ of the dynamical shell. All these possibilities 
have non-zero measurements in the phase space of $m, \; \Lambda$ and $R_{0}$, 
although the region  of gravastars is very small im comparing with 
that of black holes. Therefore, even though the existence of gravastars cannot be 
completely excluded in these dynamical models, our results do indicate that, even if
gravastars exist, they do not exclude the existence of black holes.

\begin{acknowledgments}
Part of the work was done when the authors NOS and AW were  visiting LERMA/CNRS-FRE. They 
would like to express their gratitude to the Laboratory for hospitality. This work was 
partially supported by CNPq (PR, RC, MFS,NOS and AW), FAPERJ (AM), and NSFC under grant 
No. 10775119 (AW).
  
\end{acknowledgments}

\end{document}